\documentclass[11pt]{article}
\usepackage{mathrsfs}
\usepackage[centertags]{amsmath}
\usepackage{amsfonts}
\usepackage{amssymb}
\usepackage{amsthm}
\usepackage{newlfont}
\usepackage[active]{srcltx} 
\usepackage{eufrak}
\DeclareTextFontCommand{\textfrak}{\frakfamily}
\theoremstyle{definition}
 
 \theoremstyle{remark}
 
 \numberwithin{equation}{section}

\begin{document}

\begin{center}

 \noindent {\LARGE\textbf{New MDS or near MDS self-dual codes over finite fields}}
\bigskip\bigskip

 \noindent {\Large  Hongxi Tong$^1$\ \ Xiaoqing Wang$^2$}
\medskip

 \noindent{\small  Department of
Mathematics,   Shanghai University, Shanghai 200444.$^{1\ 2}$ \\
(email:  tonghx@shu.edu.cn$^1$\ \ 2625453656@qq.com$^2$)}

\end{center}

 \noindent\textbf{Abstract:} The study of MDS  self-dual codes has attracted lots of attention in recent years.
 There are many papers on determining existence of $q-$ary MDS  self-dual codes for various lengths.
 There are not existence of $q-$ary MDS  self-dual codes of some lengths, even these lengths $< q$.
 We generalize  MDS Euclidean self-dual codes to near MDS Euclidean self-dual codes and near MDS isodual codes.
 And we obtain many new near MDS isodual codes from extended negacyclic duadic codes
 and we obtain many new MDS Euclidean self-dual codes from MDS Euclidean self-dual codes.
 We generalize  MDS Hermitian self-dual codes to near MDS Hermitian self-dual codes.
 We  obtain  near MDS Hermitian self-dual codes from extended negacyclic duadic codes
 and from MDS Hermitian self-dual codes.


\section{Introduction}

Let $\mathbb{F}_q$ denote a finite field with $q$ elements. An $[n, k, d]$ linear code $C$ over $\mathbb{F}_q$ is a $k-$dimensional subspace of $\mathbb{F}_q^n$.   the Singleton bound states a relationship among $n$, $k$ and
$d$: $d\leq n-k+1$. So the Singleton defect of a $q-ary$ linear
$[n,k,d]_q$ code $C$ is  defined by $s(C)=n-k+1-d$, $s(C)\geq 0$.

$s(C)=0$, $C$ is called an MDS code. MDS codes have very good
properties and are important. For examples Reed-Solomon codes are
MDS codes. But for an MDS code,  $n\leq k+q$ and
\begin{quotation}
Main conjecture on MDS codes$^{[6]}$: \  For a nontrivial
$[n,k,n-k+1]$ MDS code,  we have that $n\leq q+2$ if $q$ is even and
$k=3$ or $k=q-1$,  and $n\leq q+1$ otherwise.
\end{quotation}

$s(C)=1$, $C$ is called an almost MDS code.$^{[6]}$  $s(C)=s(C^{\bot})=1$,
$C$ is called a near MDS code,$^{[8]}$ where $C^{\bot}$ is the dual of $C$,
 defined as $$ C^{\bot}:=\left\{ x\in \mathbb{F}_q^n : \sum_{i=1}^n x_i y_i =0,\ \forall y \in C \right\}.$$
 Near MDS codes and almost MDS codes have many good properties as MDS codes.
There are many papers on near MDS codes and almost MDS codes.$^{[1] [2] [6] [8] [9] [19]}$

If $C$ satisfies $C=C^{\bot}$, $C$ is called Euclidean self-dual. If $C$ permutationally and monomially is equivalent to $C^{\bot}$,
$C$ is called isodual. All negacyclic self-dual codes, some well-known Hermitian self-dual and MDS codes are isodual.$^{[5]}$
And isodual codes are formally self-dual.$^{[14]}$

If $q=r^2$, the Hermitian dual code $C^{\bot H}$ of $C$ is defined as
$$C^{\bot H}:= \left\{ x\in \mathbb{F}_{r^2}^n : \sum_{i=1}^n x_i y_i^r =0,\ \forall y \in C \right\}.$$

If $C=C^{\bot H}$, $C$ is called  Hermitian self-dual.  There are many papers discussing  Hermitian self-dual codes.$^{[7] [16] [18] [20]}$
If $C$ is MDS and  Euclidean self-dual or Hermitian self-dual, $C$ is called an MDS Euclidean self-dual code or
an MDS Hermitian self-dual code, respectively.
In recent years, study of MDS self-dual codes has attracted a lot of attention.$^{[1] [10] [11] [12] [13] [15] [16] [17] [18]}$
 One of these problems in this topic is to determine existence of MDS  self-dual codes.
 When $2|q$, Grassl and Gulliver completely solve the existence of MDS Euclidean  self-dual codes in [11].
 In [12], Guenda obtain some new MDS Euclidean self-dual codes and MDS Hermitian self-dual codes.
 In [15], Jin and Xing obtain some new MDS Euclidean self-dual codes from generalized Reed-Solomon codes.
 In [18], Tong obtain many new MDS Euclidean self-dual codes from extended cyclic  duadic codes and new MDS Hermitian self-dual codes from
  generalized Reed-Solomon codes or constacyclic codes.
 But there are many MDS self-dual codes are not existence.
 For examples, a $[12, 6, 7]$ MDS self-dual code over $\mathbb{F}_{13}$ is not existence.$^{[11]}$
 There is not existence of a $[4, 2, 3]$ MDS Hermitian self-dual code over $\mathbb{F}_4$,$^{[14]}$
 and there is no MDS Hermitian self-dual code $[8, 4, 5]$ over $\mathbb{F}_{16}$.$^{[11]}$

 In this paper, we generalize these notations of MDS self-dual codes. If $C$ is  near  MDS  and  isodual,
we call $C$ a near MDS  isodual code. If $C$ is a near MDS code and  Hermitian self-dual,
we call $C$ a near MDS Hermitian self-dual code. And we  obtain them from extended negacyclic duadic codes.
We also obtain near MDS Euclidean self-dual codes, which are near MDS and  Euclidean self-dual,  by deleting some coordinates of MDS self-dual codes.
And we obtain near MDS Hermitian self-dual codes,
by deleting some coordinates of MDS  Hermitian self-dual codes.

\section{Preliminaries}

Let $(n, q)=1$ and $q$ be an odd prime power.
The negacyclic code $C$ over $\mathbb{F}_q$ of length $n$ can be considered as an ideal, $<g(x)>$, of $R_n = \frac{\mathbb{F}_q [x]}{x^n+1}$.
Let $$O_{2 n}=\left\{ 1+2 i | i=0, 1, 2, \cdots, n-1 \right\}.$$
Then $\delta^j$s ($j\in O_{2 n}$) are all solutions of $x^n+1=0$ over $\mathbb{F}_q$,
where $\delta$ is a primitive $2 n$th root of unity in some extension field $F$ of $\mathbb{F}_q$.
 The set $T \subseteq O_{2 n}$ is called the defining set of $C$, if $$T=\{j,\ j\in O_{2 n}\ \mbox{and}\ g(\delta^j)=0  \}.$$
 Obviously, the dimension of $C$ is $n-|T|$, and there is a constacyclic BCH bound on the minimum distance  of $C$,
 which states that if $T$ has $d-1$ consecutive odd integers, the minimum distance of $C$ is at least $d$.$^{[3] [4]}$

Let $a \in \mathbb{F}_q^n$. Define the discrete Fourier transform (DFT) of $a$ to be the vector $[A_0, A_1, \cdots, A_n]\in F^n$, where
$$A_i = \sum_{j=0}^{n-1} a_j \delta^{(1+2 i)j}, \ \ 0 \leq i \leq n-1.$$
And $A_i= a(\delta^{(1+2 i)})$, where $\mbox{ord} \delta =2 n$. Define
$$ A(z)=\sum_{i=0}^{n-1} A_i z^i. $$

\textbf{Lemma 1}$^{[4]}$ Let $$\theta : R_n \rightarrow F^n $$
be the negacyclic DFT map defined by $\theta(a(x))=[A_0, A_1, \cdots, A_{n-1}]$. Suppose $a(x), b(x) \in R_n$. Then

(1)\  $\theta$ is a ring homorphism.

(2)\  $A_i^q=A_{(q i + \frac{q-1}{2})}$.

(3)\  If $0 \leq t \leq n-1$, then
$$a_t= \frac{1}{n}\delta^{-t} \sum_{i=0}^{n-1} A_i \zeta^{-i t}= \frac{1}{n} \delta^{-t} A(\zeta^{-t}),$$
where $\zeta= \delta^2$.

(4)\  $\sum_{t=0}^{n-1} a_t b_t = \frac{1}{n} \sum_{i=0}^{n-1} A_i B_{-i-1}$.

(All subscripts are calculated modulo $n$.)

\textbf{Definition 1}$^{[4]}$ A $q-$splitting of $n$ is a multiplier $\mu_s$ of $n$ that induces a partition of $O_{2 n}$ such that

(1)\ $O_{2 n}= A \cup B \cup X$.

(2)\ $A$, $B$ and $X$ are unions of $q-$clotomic cosets.

(3)\ $\mu_s(A)=B$, $\mu_s(B)=A$ and $\mu_s(X)=X$.

A $q-$splitting is of Type $I$ if $X=\emptyset$. A $q-$splitting is of Type $II$ if $X=\left\{\frac{n}{2}, \frac{3 n}{2}\right\}$.

\section{Euclidean isodual Codes}

First we consider near MDS isodual codes.

\textbf{Lemma 2}$^{[4]}$  If $p$, $q$ are distinct odd primes, $q\equiv -1 (\bmod 4)$, and $r$ is the order of $q$ modulo $2 p^t$, then

(1) $\mu_{-1}$ gives
a splitting of $2 p^t$ of Type $II$ if and only if $r \not \equiv 2 (\bmod 4)$, in which case
$$x^{2 p^t}+1=\lambda A(x) \widetilde{A}(x)(x^2+1)$$
for some $\lambda \in \mathbb{F}_q$, $A(x)\in \mathbb{F}_[x]$, where $\widetilde{A}(x)= A(x^{-1}) (\bmod x^n+1)$.

(2) $\mu_{2 p^t+1}$ gives a splitting of $2 p^t$ of Type $II$ if and only if $r$ is even, in which case
$$x^{2 p^t}+1= \lambda A(x) A(-x)(x^2+1)$$
for some $\lambda \in \mathbb{F}_q$, $A(x)\in \mathbb{F}_[x]$.

\textbf{Lemma 3}$^{[4]}$  Let $q \equiv 3 (\bmod 4)$, $n=2 p_1^{e_1}\cdots p_t^{e_t}$, where $p_i$s are distinct odd primes, and let $a_i$ be an integer
that gives a splitting of $2 p_i^{e_i}$. Then $n$ has a splitting of Type $II$. Moreover, this splitting is given by $\mu_a$,
where $a$ is the unique integer in $O_{2 n}$ such that $a \equiv a_i (\bmod 2 p_i^{e_i})$.

\textbf{Theorem 1} Let $q \equiv 3 (\bmod 4)$ and $n=2 p_1^{e_1} \cdots p_t^{e_t}$, where $p_i$ are distinct odd primes. And  $r= \mbox{ord}_n q$.

 (1) $\mu_{-1}$ gives a splitting of $n$ of Type $II$ if and only if
$r \not \equiv 2(\bmod n)$.

(2) $\mu_{n+1}$ gives a splitting of $n$ of Type $II$ if and only if $r$ is even.

Proof\ \ (1) \ ($\Rightarrow$) By Lemma 2, $\mu_{-1}$ gives a splitting of $n$ of Type $II$, then $\mu_{-1}$ gives
a splitting of $2 p_i^{e_i} (1\leq i\leq t)$ of type $II$. So $r_i (= \mbox{ord}_{2 p_i^{e_i}} q) \not \equiv 2 (\bmod 4)$.
$r_i = \mbox{lcm} \left[\mbox{ord}_2 q=1, \mbox{ord}_{ p_i^{e_i}} q \right]=\mbox{ord}_{p_i^{e_i}} q$.
So $$r=\mbox{ord}_n q=\mbox{lcm}[1, r_1, r_2, \cdots, r_t]\not \equiv 2(\bmod 4).$$

($\Leftarrow$) Let $r_i=\mbox{ord}_{2 p_i^{e_i}} q \ (1\leq i \leq t)$, $q^r \equiv 1(\bmod n)$. Then $q^r \equiv 1 (\bmod 2 p_i^{e_i})$. So $r_i | r$.

If $2 \nmid r$, $2 \nmid r_i$.

If $4 | r$, $n| q^r-1$. $n \nmid q^{\frac{r}{2}}-1$ and $n \mid q^{\frac{r}{2}}+1$.  If $r_i\equiv 2(\bmod 4)$. $r_i \mid r$, so $r_i \mid \frac{r}{2}$.
$$2 p_i^{e_i} \mid  q^{\frac{r}{2}}-1, \ \mbox{and} \ 2 p_i^{e_i} \mid  q^{\frac{r}{2}}+1.$$
But it is impossible, because $(q^{\frac{r}{2}}-1, q^{\frac{r}{2}}+1)=2$ and $p_i\geq 3$.

So $$r_i \not \equiv 2 (\bmod 4), \ \ i=1, 2, \cdots, t.$$
$\mu_{-1}$ gives the splitting of $2 p_i^{e_i}$ of type $II$ by Lemma 2. By Lemma 3, $\mu_{-1}$ gives the splitting of $n$ of Type $II$.

We can prove (2) similarly by Lemma 3 and Lemma 2 (2).

\textbf{Lemma 4}$^{[18]}$  (1) Let $q \equiv 3 (\bmod 4)$ and $n=2 p_1^{e_1} \cdots p_s^{e_s} p_{s+1}^{e_{s+1}}\cdots p_t^{e_t}$, where
$$p_1 \equiv \cdots \equiv p_s \equiv 3 (\bmod 4), \  p_{s+1} \equiv \cdots \equiv p_t \equiv 1 (\bmod 4).$$
Then the equation,  $ 2 + \gamma^2 n=0$, has a solution in $\mathbb{F}_q$ if and only if $\sum_{i=1}^s e_i$ is odd.

(2) Let $q \equiv 1 (\bmod 4)$ and $n=2 n'$, where $n'$ is odd.
Then the equation,  $ 2 + \gamma^2 n=0$, has a solution in $\mathbb{F}_q$.

Let $c=(c_0, c_1, \cdots, c_{n-1})\in \mathbb{F}_q^n$, define
$$\widetilde{c}=(c_0, c_1, \cdots, c_{n-1}, c_{\infty}, c_*)\in \mathbb{F}_q^{n+2},$$
where
$$c_{\infty}= \gamma \sum_{i=0}^{\frac{n-1}{2}}(-1)^i a_{2 i}, \ \ a_* = \gamma \sum_{i=0}^{\frac{n-1}{2}}(-1)^i a_{2 i+1}.$$

Let $C\subseteq \mathbb{F}_q^n$, then $\widetilde{C} (\subseteq \mathbb{F}_q^{n+2})$ is defined to be the set $\{\widetilde{c}, c\in C\}$.

\textbf{Lemma 5}$^{[4]}$ Suppose $q$ is a prime power such that $\frac{- 2}{n}= \gamma^2$ for some $\gamma \in \mathbb{F}^*_q$,
and suppose that $D_1$, $D_2$ are odd-like negacyclic duadic codes with multiplier $\mu_s$ of Type $II$.

(1) If $s=2 n-1$, then $\widetilde{D}_i$ is self-dual for $i=1, 2$.

(2) If $\mu_{-1}(D_i)=D_i$ for $i=1, 2$, then $\widetilde{D}_1^{\bot}=\widetilde{D}_2$ and $\widetilde{D}_2^{\bot}=\widetilde{D}_1$.

\textbf{Theorem 2}  Let $q\equiv 1 (\bmod 4)$ (or $q\equiv 3 (\bmod 4)$) and $n=2 n'$, where $n'$ is odd,  and $2 n \mid q-1$ (or $2 n | q+1$).
$D_1$ and $D_2$ are negacyclic  codes with defining set
$$ T_1= \left\{1 +2 j \mid -\frac{n-2}{4} \leq j \leq \frac{n-6}{4} \right\}$$
and $$T_2= \left\{1+ 2 j \mid \frac{n+2}{4}\leq j \leq \frac{3 n-6}{4}\right\},$$ respectively.
Then $\widetilde{D}_1$  and $\widetilde{D}_2$ are  $\left[ n+2, \frac{n}{2}+1, d \geq \frac{n}{2}+1\right]$
(near) MDS isodual codes which are   extended negacyclic codes.

Proof\ \ By  definitions of $T_1$ and $T_2$
$$T_1 \cap T_2=\emptyset\ \mbox{and} \ O_{2 n}= T_1 \cup T_2 \cup \left\{\frac{n}{2}, \frac{3 n}{2}\right\}.$$
\begin{eqnarray*}
  (-1)(1+2 j) &\equiv & 1+ 2(n-1-j)(\bmod 2 n) \\
  (n+1)(1+2 j) &\equiv & 1 + 2 \left(\frac{n}{2}+j \right) (\bmod 2 n)
 \end{eqnarray*}
 So $$(-1)T_i= T_i,\  (n+1)T_i=T_{i+1 (\bmod 2)}, \ i=1, 2.$$

Case 1. When $q \equiv 1 (\bmod 4)$ and $2n|q-1$.
$$C_q (1+2 j)=1+2 j.$$

By the constacyclic BCH bound, $D_1$ and $D_2$ are $\left[n, \frac{n}{2}+1, \frac{n}{2}\right]$ MDS odd-like negacyclic codes.
Let $a=(a_0, a_1, \cdots, a_{n-1}) \in D_1$ and $\mbox{wt}(a)=\frac{n}{2}$.
$$a(x)=a_0 +a_1 x+ \cdots + a_{n-1} x^{n-1}=\alpha_1 (x^2)+ x \alpha_2 (x^2).$$
Then
$$a(\delta^{\frac{n}{2}})= \gamma^{-1} a_{\infty} + \delta^{\frac{n}{2}} \gamma^{-1} a_*.$$
If $a_{\infty}=a_*=0$, $a(\delta^{\frac{n}{2}})=0$. Then $\mbox{wt}(a)\geq \frac{n}{2}+1$.
So $\mbox{wt}(\widetilde{a}) \geq \frac{n}{2}+1$. $D_1$ is an $\left[ n+2, \frac{n}{2}+1, d \geq \frac{n}{2}+1\right]$ code.
Similarly, $D_2$  is also an $\left[ n+2, \frac{n}{2}+1, d \geq \frac{n}{2}+1\right]$ code.

$$\mu_{n+1}((a_0, a_1, a_2, \cdots, a_{n-1}, a_{\infty}, a_{*}))=(a_0, -a_1, a_2, \cdots, -a_{n-1}, a_{\infty}, -a_{*}).$$
So $$\mu_{n+1}(\widetilde{D}_i) = \widetilde{D}_{i+1 (\bmod 2)}.$$
$\widetilde{D}_1$ permutationally and monomially is equivalent to $\widetilde{D}_2$.
By Lemma 5 (2), $\widetilde{D}_1^{\bot}=\widetilde{D}_2$ and $\widetilde{D}_2^{\bot}=\widetilde{D}_1$.
So $\widetilde{D}_1$  and $\widetilde{D}_2$ are  $\left[ n+2, \frac{n}{2}+1, d \geq \frac{n}{2}+1\right]$
(near) MDS isodual codes which are   extended negacyclic codes.

Case 2. When $q \equiv 3 (\bmod 4)$ and $2 n | q+1$.
$$C_q (1+ 2 j)=-1-2 j.$$
By the constacyclic BCH bound, $D_1$ and $D_2$ are $\left[n, \frac{n}{2}+1, \frac{n}{2}\right]$ MDS odd-like negacyclic codes.
The proof can proceed as in the first case.

Next we construct (near) MDS self-dual codes from MDS self-dual codes.

\textbf{Lemma 6}$^{[11]}$ For every odd prime power $q$, there exists a self-dual MDS code of length $q+1$ over $\mathbb{F}_q$.

\textbf{Theorem 3} Assume that $q$ is a power of an odd prime such that $q \equiv 1 (\bmod 4)$.
There is a MDS Euclidean self-dual  code $C$ over $\mathbb{F}_q$ of length $2n$.
Then there is a (near) MDS  Euclidean self-dual  code $C$ over $\mathbb{F}_q$ of length $2n-2$.

Proof\ \  Let $G$ be a generator matrix of $C$, Without loss of generality, we may assume that $$G=(I_n | A)=(e_i | \alpha_i),$$
where $e_i$ and $\alpha_i$ are the rows of $I_n$(= the identity matrix) and $A$, respectively, for $1\leq i\leq n$.

We note that $$wt(\alpha_i)=n,\ \ \alpha_i \cdot \alpha_j=0,\ \ \ \alpha_i \cdot \alpha_i=-1, \ \ 1\leq i\neq j \leq n.$$

Let $c \in \mathbb{F}_q$ such that  $c^2 =-1$ ($q \equiv 1 (\bmod 4)$). $C$ has the following generator matrix:
$$G_1=\left(
              \begin{array}{c|c}
                e_1-c e_2 & \alpha_1-c \alpha_2 \\
                e_2 & \alpha_2 \\
                e_3 & \alpha_3 \\
                \vdots & \vdots \\
                e_n & \alpha_n \\
              \end{array}
            \right).$$
Deleting the first two columns and the second row of $G_1$ produces an $(n-1) \times (2n -2 )$ matrix
$$G_2=\left(
        \begin{array}{ccc|c}
          0 & \cdots & 0 & \alpha_1- c \alpha_2 \\
          \  & \  & \  & \alpha_3 \\
          \  & I_{n-2} & \  & \vdots \\
          \  & \  & \  & \alpha_n \\
        \end{array}
      \right).$$

We claim that $G_2$ is a generator matrix of some $[2n-2, n-1, d\geq n-1]$ near MDS Euclidean self-dual code $C_2$.

Obviously, the dimension of $C$ is $n-1$. And
\begin{eqnarray*}
  (\alpha_1- c \alpha_2)\cdot (\alpha_1-c \alpha_2) &=& -(c^2+1)=0, \\
  (\alpha_1- c \alpha_2)\cdot \alpha_{i+1} &=& \alpha_1 \cdot \alpha_{i+1} -c \alpha_2 \cdot \alpha_{i+1}=0, \  2 \leq i\leq n-1,\\
  1+ \alpha_{i+1}\cdot \alpha_{i+1}&=& 0, \   2 \leq i\leq n-1,\\
  0 + \alpha_{i+1} \cdot \alpha_{j+1}&=& 0, \ 2 \leq i \neq j \leq n-1.
\end{eqnarray*}
\begin{eqnarray*}
 \mbox{ wt}(\alpha_1-c \alpha_2) &\geq & n+1-2=n-1, \\
  \mbox{ wt}\left(k_1 (\alpha_1-c \alpha_2)+ \sum_{i=2}^{n-1} k_{i+1} \alpha_{i+1}\right) &\geq &
  \left\{    \begin{array}{cc} n+1- |T|, & k_1=0, \\
                                n+1-|T|-2=n-1-|T|, & k_1\neq 0, \\
              \end{array} \right.
\end{eqnarray*}
where  $$T=\{k_{i+1} | k_{i+1} \neq 0, 2 \leq i \leq n-1 \}.$$
So the minimum distance $d$ of $C_2$ is $\geq n-1$.
$C_2$ is a $[2n, n-1, d \geq n-1]$ (near) MDS Euclidean self-dual code.

From Lemma 6, there is a $[14, 7, 8]$ MDS Euclidean self-dual code over  $\mathbb{F}_{13}$.
By Theorem 3, we can obtain a $[12, 6, 6]$ near MDS Euclidean self-dual code over  $\mathbb{F}_{13}$.

\section{Hermitian Self-Dual Codes}

First, we consider conditions of $\mu_{-q}$ giving a $q^2-$splitting of $n$ of Type $I$ or Type $II$, where $n=2 n'$, $n'$ is odd.

\textbf{Theorem 4} Let $n=2 n'$, where $n' (>1)$ is odd.

(1) Let $q \equiv 1 (\bmod 4)$.  $\mu_{-q}$ gives a $q^2-$splitting of $n$ of Type $I$ and Type $II$.

(2) Let $q \equiv 3 (\bmod 4)$.  $\mu_{-q}$ gives a $q^2-$splitting of $n$ of  Type $II$ if and only if $p \nmid q^s+1$,
where $p$ is any odd prime divisor of $n$ and $s$ is any odd integer.

Proof\ \  Let $n=2 n'$, where $n'$ is odd. So $ \{\frac{n}{2}, \frac{3 n}{2}\}\subseteq O_{2 n}$, and
$$C_{q^2}\left(\frac{n}{2}\right)=\frac{n}{2},\ \  C_{q^2}\left(\frac{3 n}{2}\right)=\frac{3 n}{2}.$$

(1) Let $q\equiv 1 (\bmod 4)$. For some $j$ ($0\leq j\leq n-1$) and  $l$ ($l\geq 0$),
$$(-q)(1+2 j) \equiv (q^2)^l (1+2 j) (\bmod 2 n).$$
Then $$2 n| (q^{2 m l}+q) (1 +2 j) \ \mbox{and}\ 4 | (q^{2 m l}+q).$$
But $q^{2 m l}+q \equiv 1+1 \equiv 2 (\bmod 4)$. It is a contradiction.

So for any $j$ ($0\leq j\leq n-1$) and  $l$ ($l\geq 0$),
$$(-q)(1+2 j)\not \equiv (q^2)^l (1+2 j) (\bmod 2 n).$$
And $(-q)\frac{n}{2}\equiv \frac{3 n}{2} (\bmod 2n)$.
 So  $\mu_{-q}$ gives a $q^2-$splitting of $n$ of Type $I$ and Type $II$.

(2) Let $q \equiv 3 (\bmod 4)$,
$$(-q) \frac{n}{2} \equiv \frac{n}{2} (\bmod 2 n) \ \mbox{and}\  (-q)\frac{3 n}{2} \equiv \frac{3 n}{2} (\bmod 2 n).$$
So $\mu_{-q}$ can not give a $q^2-$splitting of $n$ of Type $I$.

If there is an odd prime $p$,   where $p|n$, and odd integer $l$ such that $p | q^l+1$,
$$\frac{n}{2 p}\in O_{2 n}, \ \ 1+ 2 j_0 = \frac{n}{2 p}, \ \mbox{for some}\ 0 \leq j_0\leq n-1.$$
So $$2 n | (q^{l+1}+q)(1+2 j_0),$$ and
$$(-q) (1+2 j_0) \equiv (q^2)^{\frac{l+1}{2}}(1+j_0) (\bmod 2 n).$$
So $\mu_{-q}$ can not give a $q^2-$splitting of $n$ of Type $II$.

If $p \nmid q^s+1$,
where $p$ is any odd prime divisor of $n$ and $s$ is any odd integer.
\begin{eqnarray*}
  2 n | ((q^2)^{\frac{s+1}{2}}+)(1+2 j) &\Leftrightarrow& \frac{n}{2} | 1+2 j \\
    &\Leftrightarrow & 1 +2 j = \frac{n }{2 } \ \mbox{or}\  \frac{3 n}{2}.
\end{eqnarray*}
So $\mu_{-q}$ gives a $q^2-$splitting of $n$ of  Type $II$.

Similarly, we can prove the next theorem.

\textbf{Theorem 5} Let $n=2 n'$, where $n'$ is odd. $\mu_{-1}$ and $\mu_{n+1}$ give $q^2-$splittings of $n$ of Type $I$ and Type $II$.

Let $c=(c_0, c_1, \cdots, c_{n-1})\in \mathbb{F}_{q^2}^n$, define
$$\overline{c}=(c_0, c_1, \cdots, c_{n-1}, c_{\infty}, c_*)\in \mathbb{F}_{q^2}^{n+2},$$
where
$$c_{\infty}= \gamma \sum_{i=0}^{\frac{n-1}{2}}(-1)^i c_{2 i}, \ \ c_* = \gamma \sum_{i=0}^{\frac{n-1}{2}}(-1)^i c_{2 i+1},$$
and $\gamma$ is a solution of equation $2+ \gamma^{q+1} n=0$ in $\mathbb{F}_{q^2}$.
Note that the equation, $2+ \gamma^{q+1} n=0$, always has a solution in $\mathbb{F}_{q^2}$.

Let $C\subseteq \mathbb{F}_{q^2}^n$, then $\overline{C} (\subseteq \mathbb{F}_{q^2}^{n+2})$ is defined to be the set $\{\widetilde{c}, c\in C\}$.

\textbf{Theorem 6} Let $n=2 n'$, where $n'$ is odd. Suppose that $D_1$, $D_2$ are odd-like negacyclic duadic
codes of length $n$ over $\mathbb{F}_{q^2}$ with multiplier $\mu_{-q}$ of Type $II$.

(1) $\overline{D}_i$ is Hermitian self-dual for $i=1, 2$.

(2) If $\mu_{-q}(D_i)=D_i$ for $i=1, 2$, then $\overline{D}_1^{\bot H}=\overline{D}_2$ and $\overline{D}_2^{\bot H}=\overline{D}_1$.

Proof\ \ (1) Let $\overline{a}, \overline{b}\in D_i$. Let $\omega=\delta^{\frac{n}{2}}$, a primitive $4$th root of unity. Define
\begin{eqnarray*}
  a(x)&=& a_0 + a_1 x + a_2 x^2 + \cdots + a_{n-1} x^{n-1}= \alpha_1(x^2) + x \alpha_2(x^2), \\
  b(x) &=&  b_0 + b_1 x + b_2 x^2 + \cdots + b_{n-1} x^{n-1}= \beta_1(x^2) + x \beta_2(x^2).
\end{eqnarray*}
So
\begin{equation*}
    a_{\infty}= \gamma \alpha_1 (-1), \  a_*=\gamma \alpha_2(-1), \ \ b_{\infty}= \gamma \beta_1 (-1), \ b_*= \gamma \beta_2 (-1).
\end{equation*}
\begin{eqnarray*}
  \sum_{t=0}^{n-1} a_t b_t^q &=& \sum_{t=0}^{n-1}\left( \frac{1}{n} \sum_{i=0}^{n-1} A_i \delta^{-(1+2 i)t}\right)
                                                 \left(\frac{1}{n^q} \sum_{j=0}^{n-1} B_j^q \delta^{-(1+2 j)q t} \right) \\
   &=& \sum_{t=0}^{n-1}\left( \frac{1}{n^{q+1}} \sum_{i=0}^{n-1} \sum_{j=0}^{n-1}  A_i B_j^q \delta^{-t[(1+2 i)+(1+2 j)q]}\right)\\
   &=& \frac{1}{n^q}\sum_{j=0}^{n-1}A_{\left(-\frac{q+1}{2}- q j \right)} B_j^q \ \ \  \
    \ \ \ \ \ \ \ \  \left( 1+2 \left(-\frac{q+1}{2}- q j\right)=(-q)(1+2 j) \right)\end{eqnarray*}
                \begin{eqnarray*}
   &=& \left\{ \begin{array}{cc}
                  \frac{1}{n^q}\left[ B^q_{\frac{n-2}{4}}A_{\frac{3 n-2}{4}}+A_{\frac{n-2}{4}}B^q_{\frac{3 n-2}{4}} \right] & q \equiv 1 (\bmod 4) \\
                  \ & \ \\
                 \frac{1}{n^q}\left[ B^q_{\frac{n-2}{4}}A_{\frac{ n-2}{4}}+A_{\frac{3 n-2}{4}}B^q_{\frac{3 n-2}{4}} \right]& q \equiv 3 (\bmod 4) \\
                \end{array}   \right.\\
         &=& \left\{ \begin{array}{cc}
                  \frac{1}{n^q}\left[ b^q (\omega) a(-\omega) + a(\omega) b^q (-\omega) \right] &  q \equiv 1 (\bmod 4) \\
                  \ &\ \\
                  \frac{1}{n^q}\left[ b^q (\omega) a(\omega) + a(-\omega) b^q (-\omega) \right]&   q \equiv 3 (\bmod 4) \\
                \end{array}
                \right. \\
    &=& \frac{2}{n^q}\left[\alpha_1(-1) \beta_1^q (-1) + \alpha_2 (-1) \beta_2^q (-1) \right] \\
    &=& \frac{2}{n^q} \gamma^{-1 -q} \left[ a_{\infty} b_{\infty}^q+ a_{*} b_{*}^q \right].
\end{eqnarray*}
So
\begin{eqnarray*}
  (\overline{a}, \overline{b}) &=& \frac{2}{n^q} \gamma^{-1 -q} \left[ a_{\infty} b_{\infty}^q+ a_{*} b_{*}^q \right]
   +\left[ a_{\infty} b_{\infty}^q+ a_{*} b_{*}^q \right] \\
   &=& \left( \frac{2}{n^q} \gamma^{-1 -q}+1 \right)\left[ a_{\infty} b_{\infty}^q+ a_{*} b_{*}^q \right] \\
   &=&  \frac{1}{n^q} \gamma^{-1 -q} (2 + n^q \gamma^{q+1} ) \left[ a_{\infty} b_{\infty}^q+ a_{*} b_{*}^q \right] \\
  &=& 0.
\end{eqnarray*}
Note that $2 + n^q \gamma^{q+1}=2 + n \gamma^{q+1} $ over $\mathbb{F}_{q^2}$.

So $\overline{D}_i$ is Hermitian self-dual for $i=1, 2$.

(2) $\overline{D}_1^{\bot H}=\overline{D}_2$ and $\overline{D}_2^{\bot H}=\overline{D}_1$ can be  proved similarly as (1).

\textbf{Theorem 7}\ Let $n=2 n'$, where $n'$ is odd. Let $D$ is a  negacyclic   code with defining set
$$ T= \left\{1 +2 j \mid \frac{n+2}{4} \leq j \leq \frac{3 n-6}{4} \right\}.$$

(1) When $q \equiv 1 (\bmod 4)$. Let $n|q+1$. Then $\overline{D}$ is an $\left[ n+2, \frac{n}{2}+1, d\geq \frac{n}{2}+1  \right]$
 (near)  MDS Hermitian self-dual code which is the extended negacyclic code.

(2) When $q\equiv 3 (\bmod 4)$. Let $n|q-1$. Then $\overline{D}$ is an $\left[ n+2, \frac{n}{2}+1, d\geq \frac{n}{2}+1  \right]$
 (near) MDS Hermitian self-dual code which is the extended negacyclic code.

Proof\ \ From $n|q+1$ or $n|q-1$, we have $2n |q^2-1$. So $C_{q^2}(1+2 j)=1+2 j$
and $D$ is an $\left[ n+2, \frac{n}{2}+1,  \frac{n}{2} \right]$ MDS negacyclic code.

(1) When $q \equiv 1 (\bmod 4)$ and $n | q+1 $. Then $q+1 =l n$, where $l$ is odd.
\begin{eqnarray*}
  (-q) (1+2 j) = -q-2 q j &=&1 -(q+1)- 2 (q+1) j +2 j \\
   &\equiv & 1 + 2 j - 2 \frac{l n}{2}\\
   &\equiv& 1 +2 \left( \frac{n}{2}+j \right) (\bmod 2 n).
\end{eqnarray*}
So $$(-q)T \cap T= \emptyset,\ \  (-q)T \cup T= O_{2 n}\setminus \{\frac{n}{2}, \frac{3 n}{2}\}.$$
And $D$  is an odd-like negacyclic duadic code. By Theorem 6, $\overline{D}$ is Hermitian self-dual.
Just like the proof of Theorem 2, we can prove that $wt(\overline{D})\geq \frac{n}{2}+1$.

Because $$(-1)T=T\ \ \mbox{and}\ \ O_{2 n}= T\cup (n+1)T \cup \left\{\frac{n}{2}, \frac{3 n}{2}\right\},$$
$\widetilde{D}$ is existence by Lemma 4 (2) and $\widetilde{D}$ is isodual by Theorem 2.
By construction methods of  $\widetilde{D}$ and $\overline{D}$, $\overline{D}$ permutationally and monomially is equivalent $\widetilde{D}$.
So $\overline{D}$ is isodual.  $\overline{D}$ is a near MDS code.

 So $\overline{D}$ is an $\left[ n+2, \frac{n}{2}+1, d\geq \frac{n}{2}+1  \right]$
 near MDS Hermitian self-dual code which is the extended negacyclic code.

(2) When $q\equiv 3 (\bmod 4)$ and $n|q-1$.
$$(-q)(1+2j) \equiv 1 +\left(\frac{n}{2}-1 -j \right) (\bmod 2n).$$
So $$ (-q)T \cap T= \emptyset,\ \  (-q)T \cup T= O_{2 n}\setminus\{\frac{n}{2}, \frac{3 n}{2}\}.$$
$D$  is an odd-like negacyclic duadic code. So the proof can proceed as in the first case.

Because the equation, $1+ c^{q+1}=0$, always has a solution in $\mathbb{F}_{q^2}$. Just like Theorem 3, we have the next theorem.

\textbf{Theorem 8} Assume that $q$ is a power of an odd prime.
There is an MDS Hermitian self-dual  code $C$ over $\mathbb{F}_{q^2}$ of length $2n$.
Then there is a near  MDS  Hermitian self-dual  code $C$ over $\mathbb{F}_{q^2}$ of length $2n-2$.

Proof\ \ Because $C^{\bot H}= (C^q)^{\bot}$, where $C^q:=\{c^q=(c_0^q, \cdots, c_{n-1}^q), \ c\in C\}.$ 
And $C^q$ and $C$ have same weighted distributions. A Hermitian self-dual  code $C$ is formally self-dual.  
Just like the proof of Theorem 3, we can prove the theorem.

\end{document}